\documentclass[twocolumn,pre,superscriptaddress,showpacs,tightenlines]{revtex4}
\usepackage{epsfig,graphicx,times}
\usepackage{amstext}
\usepackage{amsmath}
\usepackage{amssymb}
\usepackage{graphicx}
\usepackage{latexsym}
\usepackage{bm}

\def \dbarit {{\mathchar'26\mkern-11mud}}

\begin{document}

\title{Microscopic Work Distribution of Small System in Quantum Isothermal
Process}
\author{H. T. Quan}
\affiliation{Theoretical Division, MS B213, Los Alamos National Laboratory, Los Alamos,
NM, 87545, U.S.A.}
\author{S. Yang}
\affiliation{Institute of Theoretical Physics, Chinese Academy of Sciences, Beijing,
100080, China}
\author{C. P. Sun}
\affiliation{Institute of Theoretical Physics, Chinese Academy of Sciences, Beijing,
100080, China}

\begin{abstract}
For a two-level quantum mechanical system, we derive microscopically
the exact expression for the fluctuation of microscopic work in a multi-step non-equilibrium process, and we
rigorously prove that in an isothermal process, the fluctuation is
vanishingly small, and the most probabilistic work just equals to
the difference of the free energy. Our study demonstrates that the convergence of the
microscopic work in the isothermal process is due to the nature
of isothermal process rather than usual thermodynamic limit
condition. Our investigation justifies the validity of ``minimum
work principle" formulation of the second law even for a small
system far from thermodynamic limit.
\end{abstract}

\maketitle

\section{INTRODUCTION}
Thermodynamics usually deals with the systems of
infinite number of degree of freedoms, in which relative fluctuations of the
observable, e.g., energy, particle number, are inversely proportional to the
square root of the numbers of the particles of the system \cite{kersonhuang}%
. Hence for a macroscopic system consisting of infinite number of
particles, the fluctuations are vanishingly small and the ensemble
average can describe thermodynamic phenomena completely. However,
concerning small systems, usually the fluctuations of the
microscopic values of thermodynamic observable will become
appreciable, and ensemble average alone can not longer give a
complete description \cite{smallsystem}. In recent years, increasing
interests are drawn to the study of thermodynamics of small system,
and the emphases are put on the fluctuations of the microscopic
value of the observable, instead of their ensemble average. Some
notable progresses have been made, examples including the Jarzynski
equality \cite{JE, crooks} and the Fluctuation Theorem
\cite{evans}. The former connects the free energy difference of two
equilibrium states with ensemble average of microscopic work in
non-equilibrium process while the later illustrates the
probabilistic \textquotedblleft entropy decrease" of a closed system
within short time, or transient ``violation" of the second law. These studies shed new light on the understanding
of non-equilibrium thermodynamical processes of biological motors in cells and promise
important applications to the design of small-size machines. In all
these studies, for small systems, though fluctuations of most observables are appreciable, there exists an exception
-- the work done during a slowest reversible
equilibrium process (we use isothermal processes to replace slowest
reversible processes hereafter). It has been pointed out that the
fluctuation of microscopic work done by or on a small system during
a slowest reversible process is vanishingly small \cite{JE,kawai}. 
Nevertheless, though the fluctuations of microscopic work for small
systems in finite-time irreversible processes has been extensively
studied \cite{microscopic work}, and the vanishing fluctuation of microscopic work of classical small systems specially concerning
thermodynamic isothermal process has been point out, to our best knowledge, a rigorous proof of the above result from microscopic aspect is still lacking, and its quantum mechanical generalization 
has not been studied yet.

In this paper, we will investigate this problem by simulating a
quantum isothermal process with infinite number of infinitesimal
quantum adiabatic process (QAP) and quantum isochoric process (QIP)
\cite{arnaud, kieu06, quan5}. We prove rigorously from microscopic aspect the above result that, for a
two-level system, the fluctuations of the microscopic work during an
quantum isothermal process \cite{quan5} is vanishingly small. We
emphasize that, different from most cases in conventional
statistical mechanics, where fluctuations vanishes in the
thermodynamic limit, the vanishing work fluctuations for a small
system in an isothermal process is due to the intrinsic nature of
isothermal process. Our study also verify the universal
validity of the \textquotedblleft minimum work principle"
formulation of the second law: it holds even for a small system!

\begin{figure}[ht]
\begin{center}
\includegraphics[width=8cm, clip]{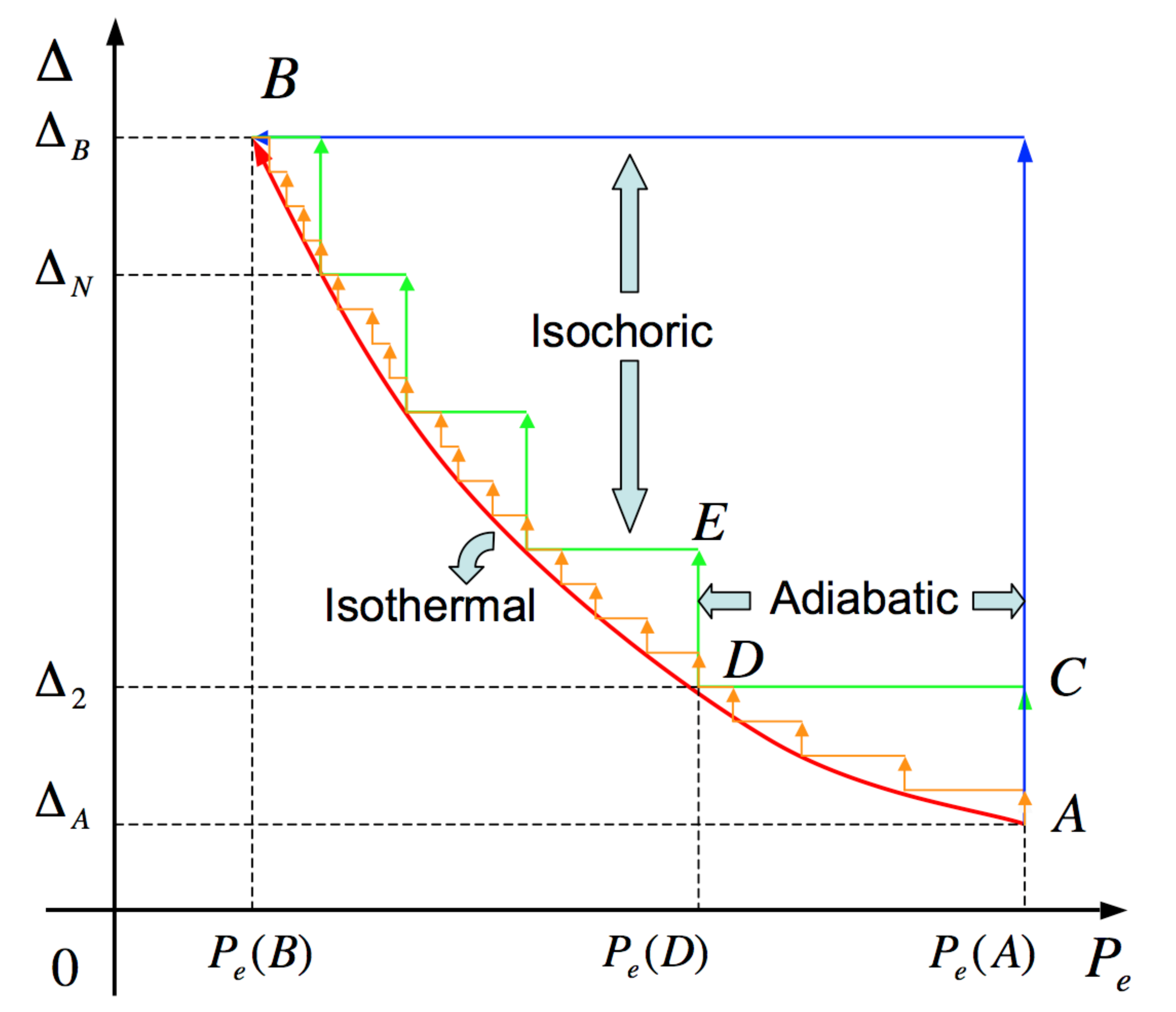}
\end{center}
\caption{Schematic illustration of a quantum isothermal process \protect\cite%
{quan5} $\widehat {AB}$. Here the horizontal axis $P_e$ is the occupation
probability in the excited state of the two-level system, and the vertical
axis indicates the level spacing of the two-level system. The smooth curve $%
\widehat {AB}$ represents the isothermal process, whose ``equation of state"
can be expressed as $\Delta (t)=-\protect\beta^{-1}\ln(P_e^{-1}-1)$. The horizontal and vertical lines represent QIC and QAP \protect\cite%
{quan5}. We can
use many small QAP and QIP to model the quantum isothermal process. For example, we use a
``five-step stair" path (green) $A\longrightarrow C \longrightarrow D \cdots \longrightarrow
B $ to simulate the smooth curve $\widehat {AB}$. ``One-step" path (blue) and ``twenty-step" path (orange) are also illustrated. }
\label{fig1}
\end{figure}

\section{The thermodynamic process in parameter space}

We consider a two-level
quantum mechanical system with excited (ground) states $\left \vert
e\right
\rangle $ ($\left \vert g\right \rangle $) with instantaneous
eigen-energy $E_{e}(t)$ ($E_{g}(t)$) depending on time $t$. This two-level system can be modeled as a spin-1/2 in an external magnetic field. It interacts
with a heat bath of inverse temperature $\beta$, which can be universely
modeled as a collection of many bosons with creation (annihilation)
operators $a_{q}^{\dag }$ ( $a_{q})$ \cite{leggett}. The model Hamiltonian
reads \cite{spinboson, berman}.
\begin{equation}
H=\Delta (t)\sigma _{z}+\sum_{q}\omega _{q}a_{q}^{\dag
}a_{q}+\sum_{q}(\lambda _{q}\sigma _{-}a_{q}^{\dag }+h.c.),  \label{1}
\end{equation}%
where $\sigma _{-}=\left \vert g\right \rangle \left \langle e\right \vert
=(\sigma _{x}-i\sigma _{y})/2$ and $\sigma _{z}=(\left \vert e\right \rangle
\left \langle e\right \vert -\left \vert g\right \rangle \left \langle
g\right \vert )/2$. Initially, let the two-level system be thermalized to
equilibrium. Then we alter the magnetic field slowly so that the energy
level spacing $\Delta (t)$ slowly changes from $\Delta _{A}$ to $\Delta _{B}$%
. During the controlling process illustrated by the smooth curve $\widehat{AB%
}$ in Fig. 1, the work is done on the system. In the infinitely slow
process, which can be alternatively regarded as a quantum isothermal process
\cite{quan5}, the two-level system is in the thermal equilibrium at every
instant, which is described by the diagonal reduced density matrix $\rho
_{S}(t)=P_{e}(t)\left \vert e\right \rangle \left \langle e\right \vert
+[1-P_{e}(t)]\left \vert g\right \rangle \left \langle g\right \vert $,
where $P_{e}(t)=\exp [{-\Delta (t)]/}(1+\exp [{-\beta \Delta (t)]})$
satisfies the Gibbs distribution. It should be pointed out that, during the
isothermal process, there is a heat exchange between the two-level system
and the heat bath.

For such an isothermal process, it is difficult to calculate the microscopic
work distribution directly. According to Ref. \cite{ arnaud,kieu06,quan5},
however, this process can be simulated by a series of QAP and QIP. In QAP (QIC) processes, there is only work done
(heat exchange). Hence, \ using the changes of eigen-energies of microscopic
state \ at instant $t=A,C,D$ $,$ we can indirectly calculate the microscopic
work done (heat exchange) \cite{JE, crooks} $\dbarit W=E_{\alpha
}(C)-E_{\alpha }(A)$ ($\dbarit Q=E_{\alpha }(D)-E_{\beta }(C)$), for $\alpha
,\beta =e,g$. In the parameter space, these QAP and QIP series
processes are represented by the \textquotedblleft stair" path ($%
A\longrightarrow C\longrightarrow D\longrightarrow \cdots \longrightarrow B$%
) in Fig. 1. When every step of the \textquotedblleft stair" path becomes
infinitesimal, the \textquotedblleft stair" path becomes equivalent to the
isothermal process $\widehat{AB}$. In this way we simulate the quantum
isothermal process with $N$ equal-height steps (see Fig. 1) with the small
height $\Delta =(\Delta _{B}-\Delta _{A})/N$ where $\Delta _{A}$ and $\Delta
_{B}$ are the level spacings at point $A$ and point $B$ respectively. The
level spacings of the two-level system after the $(j-1)$-th QIC is
\begin{equation}
\Delta _{j}=\Delta _{A}+(j-1)\Delta ,  \label{2.5}
\end{equation}%
for $j=1,2,\cdots ,N+1.$\ The initial and final point $A$ and $B$
corresponds to $j=1$ and $j=N+1$ respectively. When we fix the initial point
$A$, and the final point $B$, the jump $\Delta $ in every step decrease with
the increase of the step number $N$, and $\Delta $ approaches zero when $N$
becomes infinity. Obviously, when $N\longrightarrow \infty $, the
\textquotedblleft stair" path approaches its asymptotic behavior - the
isothermal path (see Fig. 1). When the system reaches thermal equilibriums, the
occupation probabilities obeys the Gibbs distribution defined by
\begin{equation}
P_{e}^{j}=e^{-\beta \Delta _{j}}[1+e^{-\beta \Delta _{j}}]^{-1} ; P_{g}^{j}=P_{e}^{j}e^{\beta \Delta _{j}}\label{2.8}
\end{equation}

We remark that there are three time scales in our process: $\tau _{a}$ for
quantum adiabatic approximation, $\tau _{c}$ the control time of changing
the magnetic field, hence the level spacing, and $\tau _{r}$ the relaxation
of the two-level system. According to Ref. \cite{berman} , $\tau _{r}$ is
determined by the coupling strength $\lambda _{q}$ (\ref{1}). We consider
the case that $\tau _{a}\ll \tau _{c}\ll \tau _{r}$ for a quantum adiabatic
process where we can define the microscopic work in every realization of the
process.

\section{Microscopic work distribution}
Having defined the \textquotedblleft
path" in the parameter space ($\Delta -P_{e}$) space, we can further
introduce the microscopic work and its corresponding probabilities for a
given \textquotedblleft path". Actually, the definition of microscopic work
is very similar to that in Ref. \cite{crooks}. In the above path divided
into many \textquotedblleft steps", the first step $A\longrightarrow
C\longrightarrow D$ consists of a QAP $%
A\longrightarrow C$, and a QIP $C\longrightarrow D$.
At the beginning (the point $A$ of Fig. 1), the system is initially in a
thermal equilibrium state $\rho _{S}(A)$, which implies that the system is
either in its microscopic state $\left \vert g\right \rangle $ or $%
\left
\vert e\right \rangle $ with probabilities $P_{g}^{1}$ and $P_{e}^{1}$
respectively. We choose the ground state in the energy reference point so
that the microscopic energy $E(A)$ of the system at initial pint $A$ can
take $E_{e}(A)=\Delta _{A}$ or $E_{g}(A)=0$, with probability $P_{e}^{1}$
and $1-P_{e}^{1}$ respectively. In the first QAP $%
A\longrightarrow C$, the system remains in its microscopic state $%
\left
\vert g\right \rangle $ ($\left \vert e\right \rangle $) if the
system is initially in its microscopic state $\left \vert g\right \rangle $ (%
$\left
\vert e\right \rangle $). As there is no heat exchange in the
QAP, the work done by external controller is just the
change of the microscopic energy $W_{\alpha }=E_{\alpha }(C)-E_{\alpha }(A)$
for $\alpha =e,g$. Correspondingly the work done during $A\longrightarrow C$
can be either $\Delta _{C}-\Delta _{A}$ or $0$ with probabilities $%
P_{e}^{1}$ or $1-P_{e}^{1}$ respectively. This also agrees with the
definition of work in quantum mechanical system: work is associated with the
change of the level spacing \cite{quan5, kieu06, quan1}.

After the QAP, a quantum isochroc process $%
C\longrightarrow D$ (see Fig. 1) follows. Here, there is no work done
according to the definition of work in quantum mechanical system \cite%
{quan5,  kieu06,quan1}, because there is no change in the eigenergies.
Nevertheless, there is heat exchange between the system and the bath. The
QIP last long enough ($\gg \tau _{r}$) so that the system can
reach thermal equilibrium with the heat bath. After a thermolization for
long-time, the two-level system reach thermal equilibrium with the heat bath
again (\ref{2.8}) at instant $D$ indicated in Fig. 1. Then a second step $%
D\longrightarrow E\longrightarrow F$ begins. Similarly, the microscopic work
$0$ or $\Delta _{3}-\Delta _{2}$ is done in this step with probabilities $1-P_{e}^{2}$ or $P_{e}^{2}$. The microscopic work done and their
probabilities for the remaining steps can be obtained through a similar
analysis. Because in every QIP, the system is independently
thermalized by the heat bath, then there should be no correlations of the
probabilities distributions in every two neighbor steps, or alternatively,
this process is Markovian process. Hence, the total microscopic work done
after the $N$-step is a sum of microscopic works done in all steps and the
joint probabilities for the $N$-step as a whole is the product of that of
all steps.

For a special example that the microscopic work done during the whole
process is $W=N\Delta $ where $\Delta $ is that for each QIP step, the
joint probabilities for the system keeping in $\left \vert e\right \rangle $
in every QIP is $P\left[ N\Delta \right] =P_{e}^{1}P_{e}^{2}%
\cdots P_{e}^{N}$. The more general case with microscopic work $W=\left(
N-k\right) \Delta $ corresponds to a microscopic process, in which $k$ out
of $N$ QIPs ends with the system in its microscopic state $%
\left \vert g\right \rangle $. The probability $P(k):=P\left[ \left(
N-k\right) \Delta \right] $ with the microscopic work $W=\left( N-k\right)
\Delta $ in the $N$-step path is given by the following \ eqution:\textbf{\ }
\begin{equation}
P(k)=\left(\prod_{j=1}^{N}P_{e}^{j}\right)\left(\prod_{l=0}^{k-1}\frac{e^{\beta \Delta
_{B}}-e^{\beta (\Delta _{A}+l\Delta )}}{e^{\beta (l+1)\Delta }-1}\right),  \label{5}
\end{equation}
To prove the above result, we first consider the case with $k=1$. For this case, there
is one and only one out of the $N$ QIPs, in which the system ends
up in the microscopic state $\left \vert g\right \rangle $. Then the
corresponding probability can be caculated as $P(1)=\left(
1-P_{e}^{1}\right) P_{e}^{2}\cdots P_{e}^{N}+P_{e}^{1}\left(
1-P_{e}^{2}\right) \cdots P_{e}^{N}+\cdots +P_{e}^{1}P_{e}^{2}\cdots \left(
1-P_{e}^{N}\right) =\left(\prod_{j=1}^NP_e^j\right)e^{\beta \Delta_B}\sum_{x_1=1}^N e^{-\beta \Delta x_1}$ or
\begin{equation}
P(1)=\left(\prod_{j=1}^{N}P_{e}^{j}\right)\frac{\left( e^{\beta \Delta _{B}}-e^{\beta
\Delta _{A}}\right)}{e^{\beta \Delta }-1},  \label {6.2}
\end{equation}%
That means thre Eq. (\ref{5}) holds for $k=1$. Similarly we can check the case with $
k=2$. For this case, there are two out of the $N$ QIP, in which the system ends
up in the microscopic state $\left \vert g\right \rangle $. Hence its probability can be expressed as $P(2)=\left(
1-P_{e}^{1}\right)\left(1-P_{e}^{2}\right) P_{e}^{3}\cdots P_{e}^{N}+\left(1-P_{e}^{1}\right)P_{e}^{2}\left(
1-P_{e}^{3}\right) \cdots P_{e}^{N}+\cdots +P_{e}^{1}P_{e}^{2} \cdots \left(1-P_{e}^{N-1}\right) \\\left(
1-P_{e}^{N}\right) =\left(\prod_{j=1}^NP_e^j\right)e^{2\beta \Delta_B}\sum_{x_1=1}^N \sum_{x_2=1}^{x_1-1} e^{-\beta \Delta (x_1+x_2)}$
or
\begin{equation}
P(2)=\left(\prod_{j=1}^{N}P_{e}^{j}\right)\frac{\left[ e^{\beta \Delta _{B}}-e^{\beta
\Delta _{A}}\right] \left[ e^{\beta \Delta _{B}}-e^{\beta
(\Delta _{A}+\Delta)}\right]}{(e^{\beta \Delta }-1)(e^{2\beta \Delta }-1)}, \label {6.3}
\end{equation}
Hence Eq. (\ref{5}) also holds for $k=2$ case.
In general, for an arbitrary $k$, the corresponding probability can be expressed as
 \begin{eqnarray}
\begin{split}
P(k)=\left(\prod_{j=1}^NP_e^j\right)e^{k\beta \Delta_B}\chi(k), \label {6.4}
\end{split}
\end{eqnarray}
where $\chi(k)=\sum_{x_1=1}^N \sum_{x_2=1}^{x_{1}-1} \cdots\sum_{x_{k}=1}^{x_{k-1}-1}e^{-\beta \Delta (x_1+\cdots+x_k)}$.
As $\chi(k), (k=1, 2, \cdots, N)$ satisfy the following relation
\begin{eqnarray}
\begin{split}
\chi(k)&=\sum_{i=1}^{k-1}\left[ \left(\prod_{j=1}^{i} \frac{-1}{1-e^{-j\Delta}}\right)\left( -e^{-i\Delta}\right)\chi(k-i)\right]\\
&+ \left(\prod_{j=1}^{k}\frac{-1}{1-e^{-j\Delta}}\right)\left[ e^{-k(N+1)\Delta}- e^{-k\Delta}\right], \label {6.6}
\end{split}
\end{eqnarray}
we can use the complete induction method to prove that the $\chi(k)$ can be generally expressed as
 \begin{eqnarray}
\begin{split}
\chi(k)=\prod_{l=0}^{k-1}\frac{e^{-\beta N\Delta }\left[ e^{\beta N\Delta}-e^{\beta l\Delta}\right] }{e^{\beta(l+1)\Delta}-1}
 \label {6.5}
\end{split}
\end{eqnarray}
Substituting Eq. (\ref {6.5}) into Eq. (\ref {6.4}), we obtain Eq. (\ref {5}). Hence, by now we prove the
general result given by Eq. (\ref{5}).

\section{Most probabilistic distribution and fluctuation}
The above equation (%
\ref{5}) can result in the main conclusion in this paper. From the above
microscopic work distribution function (\ref{5}), we obtain the ratio $%
R(k)=P(k+1)/P(k)$ of distributions for two close microscopic work, i.e.,
\begin{equation}
R(k)=\frac{\left( e^{\beta \Delta _{B}}-e^{\beta \Delta
_{A}+k\beta \Delta }\right) }{e^{\beta \left( k+1\right) \Delta }-1}.
\end{equation}%
Let $\tilde{k}$ \ maximaze the probability distribution $P(k)$ for the
microscopic work $[N-(\tilde{k}+1)]\Delta $ . Then $P(\tilde{k})\geq P(%
\tilde{k}\pm 1),$or $R(\tilde{k})\leq 1$ or $R(\tilde{k}-1)\geq 1$. For very
large $\tilde{k},R(\tilde{k})\simeq 1$ that
\begin{equation}
\tilde{k}\Delta =\frac{1}{\beta}\ln \left[ \frac{1+\exp [\beta \Delta _{B}]}{\exp [\beta (\Delta _{B}-\Delta _{A})/N]+\exp
[\beta \Delta _{A}]}\right] .
\end{equation}%
In the large $N$ limit, the above equation determines the microscopic
work $\tilde{W}=(N-\tilde{k})\Delta $ with most probabilistic distribution
\begin{equation}
\tilde{W}=\frac{1}{\beta }\ln \left( \frac{1+e^{\beta \Delta _{B}}}{%
1+e^{\beta \Delta _{A}}}\right)
\end{equation}%
which is just the free energy difference $\Delta F_{AB}=F_{B}-F_{A}$ where $%
F_{j}=\ln [1+\exp (\beta \Delta _{j})]/\beta $ for $j=A,B$
\begin{figure}[th]
\begin{center}
\includegraphics[width=4cm, clip]{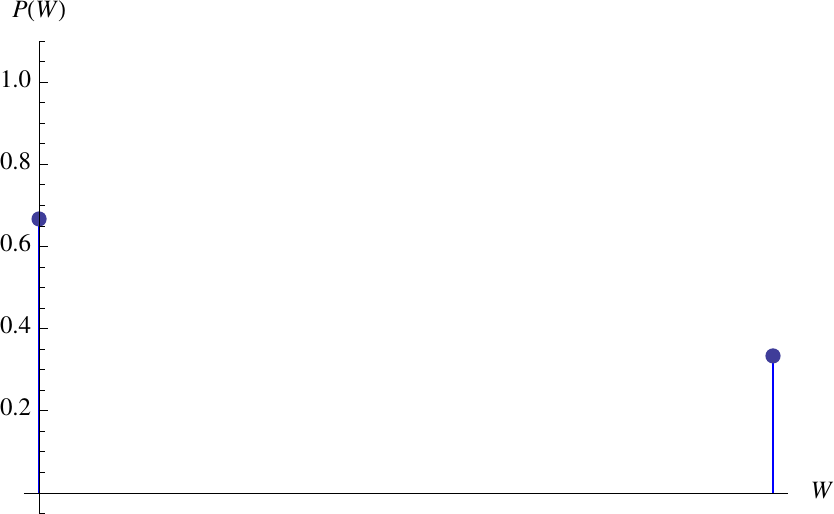}
\includegraphics[width=4cm,clip]{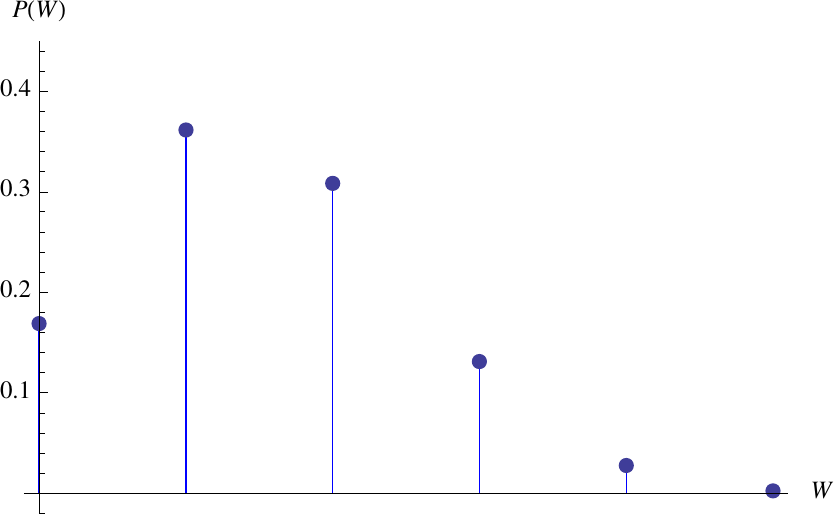}
\includegraphics[width=4cm, clip]{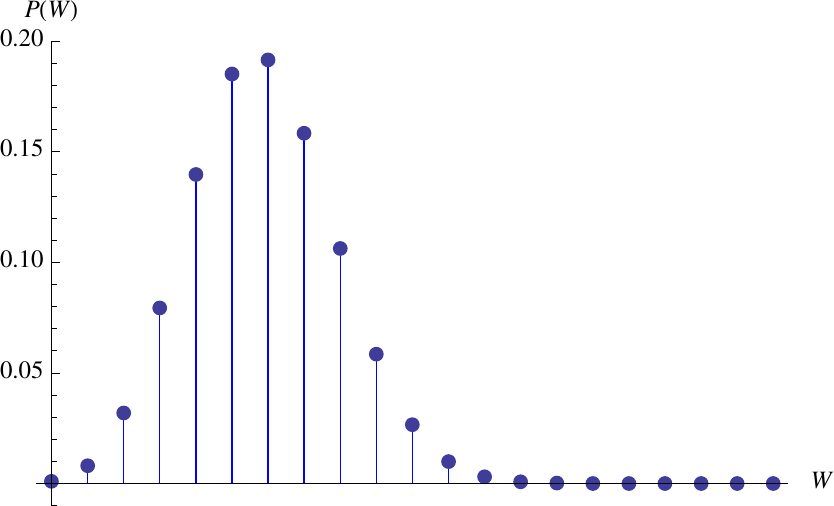} %
\includegraphics[width=4cm, clip]{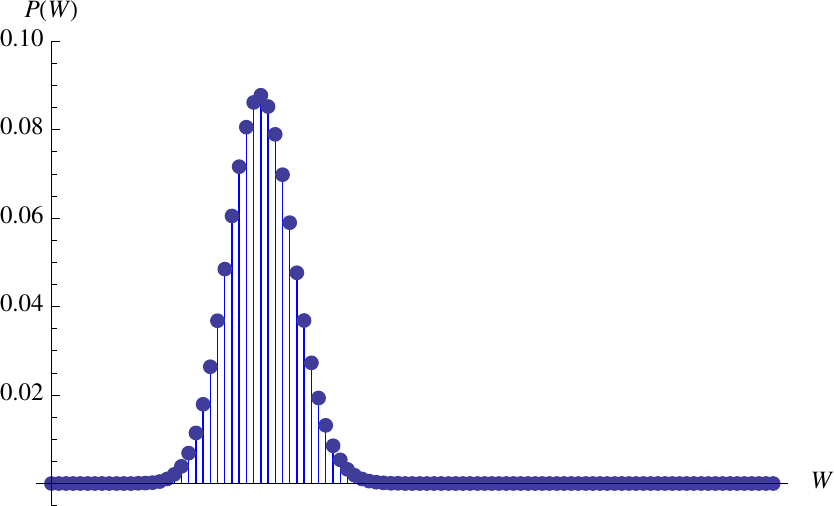}
\includegraphics[width=4cm,clip]{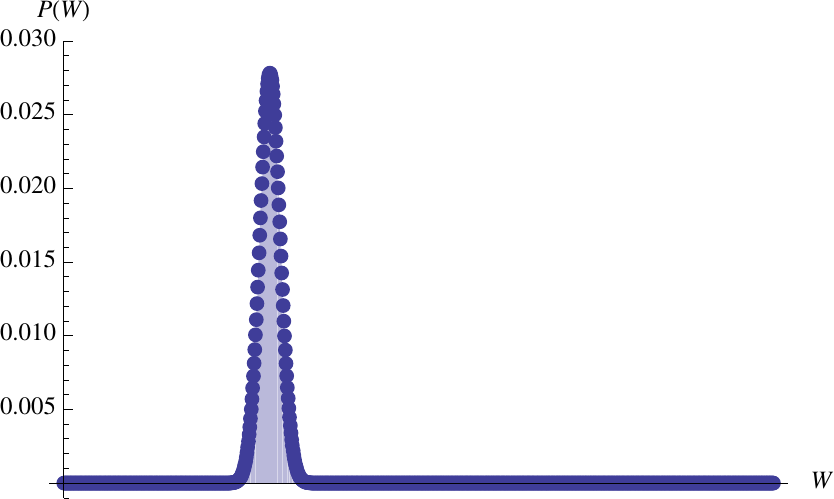}
\includegraphics[width=4cm, clip]{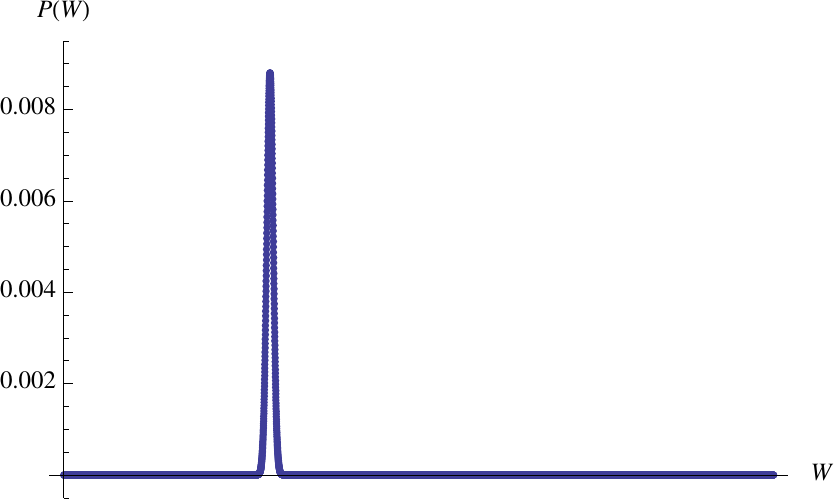}
\end{center}
\caption{Microscopic work distribution of an $N$-step \textquotedblleft
stair" process. The horizontal axis indicates the possible microscopic work
ranging from $0$ to $\Delta _{B}-\Delta _{A}$, and the vertical axis is
their probabilities. Here, $\exp (-\protect\beta \Delta _{A})=1/2$, and $%
\exp (-\protect\beta \Delta _{B})=1/3$. The steps are chosen to be $N=1$, $5$%
, $20$, $100$, $1000$, and $10000$ respectively. ``Path" corresponding to $N=1$, $5$%
, $20$ are given in Fig. 1. From these figures it can
be inferred that when $N$ is small the process is irreversible, and the
fluctuation is appreciable. The relative fluctuation of the microscopic work
vanishes when $N\rightarrow \infty $, or the fluctuation of an isothermal
process approaches zero. Besides, the most probabilistic work from the  (numerical)
figures $\tilde{W}=0.29(\ln 3-\ln 2)k_{B}T$ agrees well with the (analtical) free energy
difference $\Delta F_{AB}=[\ln (1+1/2)-\ln (1+1/3)]k_{B}T$.}
\end{figure}

Next let us give a heuristic analysis of the dispersion of the work
distribution (\ref{5}). Because all steps in the \textquotedblleft stair"
path are  independent with each other, thus the whole process can be
regarded as Markovian. So  the variance of  total microscopic work done
during the whole process equals to the sum of variance of local microscopic
work in every step, i.e.,  $\left\langle W_{AB}^{2}\right\rangle
-\left\langle W_{AB}\right\rangle ^{2}=\sum_{j=1}^{N}(\left\langle
W_{j}^{2}\right\rangle -\left\langle W_{j}\right\rangle ^{2})$, where $W_{j}$
is the microscopic work done during the $j$th QAP, and the
local fluctuations
\begin{equation}
\left\langle W_{j}^{2}\right\rangle -\left\langle W_{j}\right\rangle
^{2}=\Delta ^{2}[P_{e}^{j}-(P_{e}^{j})^{2}]
\end{equation}%
for different $j$ are  similar. Here  $\Delta $ is inversely proportional to
$N$, and $\left\langle W_{AB}\right\rangle $ being independent of $%
N$, the relative variance of $W_{AB}$ is inversely proportional to
\begin{equation}
\sqrt{\frac{\left\langle W_{AB}^{2}\right\rangle -\left\langle
W_{AB}\right\rangle ^{2}}{\left\langle W_{AB}\right\rangle }}\propto \frac{1%
}{\sqrt{N}}
\end{equation}

We numerically plot the work distribution function (see Fig.2) based
on the above analytical result (\ref{5}) to test the above analysis.
Here we choose the step number $N$ from $1$ to $10000$. For $N=1$,
the ``stair" path becomes a ``one-step" path consists of an QAP and
an QIP (see Fig. 1). The microscopic work corresponding to the
``one-step" path is either $\Delta_B-\Delta_A$ or $0$ with the
probability $P(W=\Delta_B-\Delta_A)=P_e^1$ or $P(W=0)=1-P_e^1$. In
the above figures, we choose $\exp (-\protect\beta \Delta
_{A})=1/2$, ($P_e^1=1/3$), and the numerical result agrees well with
our analysis.  For $N=5$ (see Fig. 1), the possible microscopic work
can be $W=i (\Delta_B-\Delta_A)/5, i=0,1,2, \cdots, 5$. The
numerical result indicates vanishing probability for
$W=\Delta_B-\Delta_A$. For $N=20$ (see Fig. 1), the numerical result
show even more vanishing probabilities of microscopic work. That is,
the dispersion (fluctuation) of microscopic work decrease with the
increase of $N$. Actually, from the above numerical figures, it is
not difficult to find that the dispersion of the microscopic work
distribution is inversely proportional to the square root of $N$.
For example, the dispersion for $N=100$ is ten times that for
$N=10000$ case (see Fig. 2). Hence, numerical results agrees well
with our heuristic analysis and both they verified our main result,
when $N\longrightarrow \infty$, the fluctuations of microscopic work
vanishes.

\section{Minimum work principle for a two-level system}
As we have mentioned before, for small systems and within short
time, the formulation \textquotedblleft entropy
never decrease for a closed system" of the second law may be transiently
\textquotedblleft violated" probabilistically due to appreciable fluctuations
\cite{evans}. A straightforward question is: will the other formulations of the second law, e.g., the
minimum work principle \cite{kawai,minimum}, also be transiently
\textquotedblleft violated" probabilistically for small systems? The
\textquotedblleft minimum work principle" states that
\textquotedblleft when varying the speed of a given process for an
initially equilibrium
system, the work is minimal for the slowest realization of the process" \cite
{kawai,minimum}. In the following we will test the validity of
``minimum work principle" for a two-level system by utilizing the
formula (\ref {5}) we derived above. The average work over all
possible realizations for a given $N$-step path can be expressed as
\begin{equation}
\left\langle W\right\rangle_N= \sum_{k=0}^{N} \left(\prod_{j=1}^{N}P_{e}^{j}\right) \left(\prod_{l=0}^{k-1}\frac{e^{\beta \Delta
_{B}}-e^{\beta (\Delta _{A}+l\Delta )}}{e^{\beta (l+1)\Delta }-1}\right) (N-k) \Delta, \label{15}
\end{equation}

\begin{figure}[th]
\begin{center}
\includegraphics[width=8cm, clip]{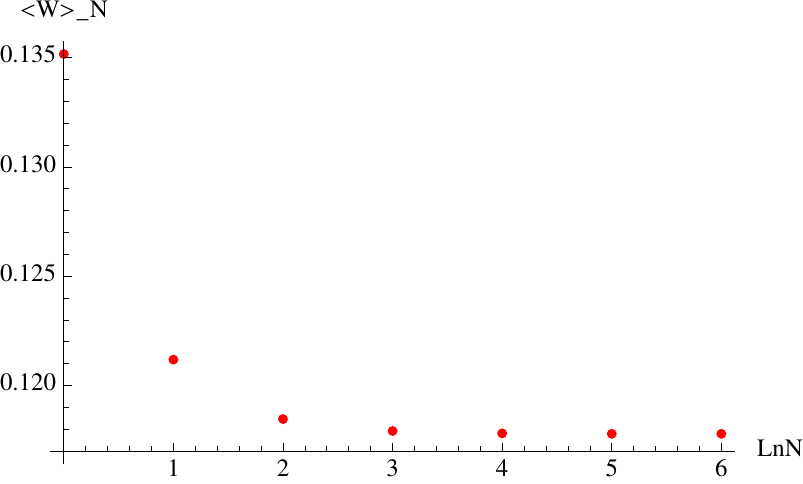}
\end{center}
\caption{Averaged work $\left\langle W\right\rangle_N$
as a function of $N$ (\ref{15}). The steps $N$ chosen here are $5^0=1, 5^1=5, 5^2=25, 5^3=125, 5^4=625, 5^5=3125$, and $5^6=15625$. It 
can be seen that the averaged work is a monotonically deceasing function of $N$.
In the one-step path ($N=1$), the averaged work equals to $\left\langle W\right\rangle_1=(\ln{3}-\ln{2})/3\approx0.135155 k_{B}T$;
In the $15625$-step path, the averaged work equals to $\left\langle W\right\rangle_{5^6}=(\ln{3}-\ln{2})/3\approx0.11784 k_{B}T$, which is very close
to its asymptotic value $\Delta F_{AB}=[\ln (1+1/2)-\ln (1+1/3)]k_{B}T
\approx0.117783 k_{B}T$. Thus, it can be inferred that the ``minimum work principle" still holds for a two-level system.}
\end{figure}

In Fig. 3 we plot the averaged work $\left\langle W\right\rangle_N$
as a function of $N$ (\ref{15}). It can be seen that for the
two-level system, $\left\langle W\right\rangle_N$ is a monotically
decreasing function of $N$ (time $t$), and when $N \rightarrow\infty, (t\rightarrow\infty)$, the
averaged work $\left\langle W\right\rangle_N$ approaches an
asymptotic value, and its minimum value -- the difference of the free energy. Thus, from the numerical result it can be inferred that the ``minimum work principle" still holds for a two-level system.

The above proof of minimum work principle can be alternatively
understood in the following way. From the above analytical and
numerical result, we observed  that the fluctuation of microscopic
work in an isothermal process vanishes, and then the work of the
most probabilistic
distribution equals to the difference of the free energy $\tilde{W}=\Delta F$%
. According to  Ref. \cite{JE},  $\left\langle W_{\mathrm{irre}%
}\right\rangle \geqslant \Delta F$, where $\left\langle W_{\mathrm{irre}%
}\right\rangle $ is the average work done during an irreversible
process.
Combining the two results, we have $\left\langle W_{\mathrm{irre}%
}\right\rangle \geqslant \tilde{W}$. Thus we proved the minimum work
principle for small system.

\section{Discussion and conclusion}

Before concluding this paper, we would like to emphasize the following points: First, the technique of simulating isothermal processes with adiabatic processes
and isochoric processes are important to our proof, which enables us to establish the connection between large time limit and large N limit.
Second, the calculation of exact expression of microscopic work in our paper is non-trivial because
the work contributions in the different steps are not identically distributed. Hence, it is different from the law of large numbers, with
time as the large number \cite{kawai}. Third, we proved the \textquotedblleft
minimum work principle" formulation of the second law stands for
even small system, though other formulations may be transiently ``violated"
probabilistically \cite{evans}. This is not surprising because ``minimum work principle" concerns infinite-long-time processes,
which has no contradiction with the transient ``violation" of the second law  for small systems predicted by the Fluctuation Theorem. Actually, the Fluctuation Theorem
does not constitute real violation of the second law, which is a statistical law and holds when averaged over different realization of the process.
Fourth, the isothermal process is reversible, but the finite $N$
\textquotedblleft step path" is irreversible, due to the QIP
(thermolization) is irreversible. We can thus expect that the work
dissipation \cite {kawai, microscopic work} for the finite $N$ step
path will be finite and will decrease with the increase of $N$, and
finally vanishes when $N$ approaches infinity.

In summary, by simulating an quantum isothermal process with infinite many infinitesimal
QAP and QIP, we obtain the analytical
expressions of microscopic work distribution in an isothermal process.
Through both analytical and numerical analysis, we rigorously verify that the fluctuations of microscopic work distribution vanishes
even for a small system in an isothermal process. This result is different
from the usual fluctuations in statistical mechanics, e.g., the energy
fluctuation and particle number fluctuation in canonical ensamble and grand
canonical ensamble, where the fluctuations of energy and particle nubmers
approaches zero when the system approaches thermodynamic limit (particle number approaches infinity $%
N_{P}\longrightarrow \infty $). Here, however, even for single particle
system, we microscopically demonstrate the vanishing of microscopic work fluctuation. Because $%
N\longrightarrow \infty $ is a must to simulate an isothermal process, we
conclude that the vanishing of microscopic work fluctuations is due to the intrinsic
nature of isothermal process, rather than the thermodynamic limit of the
system size. We also prove that for a small system, the ``minimum work principle" formulation of the second law
holds though other formulations maybe transiently ``violated" probabilistically. Finally we would like to point it out that our result is universal and does not
depend on the specific model used here, because the technique of simulating the isothermal process
with the isochoric process and the adiabatic process can be applied to any systems. Generalizations of our current discussion to other models
will be given in the future.

\section{acknowledgments}
The authors thank a anonymous referee for pointing out a mistake in our previous version of the manuscript.
HTQ thanks Rishi Sharma for stimulating discussions and gratefully acknowledges the support of the U.S. Department of Energy through
the LANL/LDRD Program for this work; CPS is supported by NSFC with grant Nos. 90203018, 10474104, 60433050, 10704023 and NFRPC with grant Nos. 2006CB921205, 2005CB724508.

\end{document}